\documentclass[graybox]{svmult}

\usepackage{mathptmx}       % selects Times Roman as basic font
\usepackage{helvet}         % selects Helvetica as sans-serif font
\usepackage{courier}        % selects Courier as typewriter font
\usepackage{type1cm}        % activate if the above 3 fonts are
\usepackage{makeidx}         % allows index generation
\usepackage{graphicx}        % standard LaTeX graphics tool
\usepackage{multicol}        % used for the two-column index
\usepackage[bottom]{footmisc}% places footnotes at page bottom
\usepackage{epsfig,color}
\usepackage{rotating}

\makeindex             % used for the subject index
\begin{document}

\institute{David Martinez-Delgado, Ramon y Cajal fellow \at Instituto de Astrofisica de Canarias, \email{ddelgado@iac.es}}
\title*{A pilot survey of stellar tidal streams in nearby spiral galaxies}
% Use \titlerunning{tidal streams} for an abbreviated version of
\author{ David Mart\'\i nez-Delgado$^{1,2}$, R. Jay Gabany$^{3}$, Jorge Pe\~narrubia$^{4}$, Hans-Walter Rix$^{2}$, Steven R. Majewski$^{5}$, Ignacio Trujillo$^{1}$and Michael Pohlen$^{6}$ \\ $^{1}$ Instituto de Astrof\'\i sica de Canarias, La Laguna, Spain \\ $^{2}$ Max-Planck-Institute fur Astronomie, Heidelberg, Germany\\$^{3}$ Black Bird Observatory, New Mexico, USA\\
$^{4}$ Institute of Astronomy, Cambridge, UK\\$^{5}$ Department of Astronomy, University of Virginia, USA\\$^{6}$Cardiff University, UK}

\authorrunning{Martinez-Delgado et al.}

\maketitle

\abstract{Within the hierarchical framework for galaxy formation, merging
and tidal interactions are expected to shape large galaxies 
to this day. While major mergers are quite rare at present,
minor mergers and satellite disruptions -  which result in stellar streams -  should be common, and are indeed seen in both the Milky Way and the Andromeda Galaxy. As a pilot study, we have carried out ultra-deep, wide-field imaging of 
some spiral galaxies in the Local Volume, 
which has revealed external views of such stellar tidal streams at unprecedented  detail, with data taken at small robotic telescopes (0.1-0.5-meter) that provide exquisite surface brightness sensitivity. The goal of this project is to undertake the first systematic and comprehensive imaging
survey of stellar tidal streams, from a sample of $\sim$ 50 nearby Milky-Way-like spiral galaxies within 15 Mpc, that features a surface brightness sensitivity of $\sim$ 30 mag/arcsec$^{2}$. The survey will result in estimates of the incidence, size/geometry and stellar luminosity/mass
distribution of such streams. This will not only put our Milky Way and M31 in context but, for the first time, also provide an extensive statistical basis for comparison with  state-of-the-art,
 self-consistent cosmological simulations of this phenomenon.}

\section{Introduction}
\label{sec:1}

Within the hierarchical framework for galaxy formation (e.g., White \& Rees 1978), 
the stellar bodies of galaxies are expected to form and evolve through dark-matter-driven 
mass in-fall and successive coalescence of smaller, distinct sub-units that span a wide mass range.
Mergers of initially bound sub-halos (which we refer to as {\it satellites}; they consist of dark matter,
gas, and in most cases stars) are effected by dynamic
friction, either through gradual orbital decay or by a single encounter
(depending on the initial orbit), its eccentricity and the
satellite-to-main-galaxy mass ratio. It is likely that a satellite becomes
disrupted by the tidal forces of the larger companion before its orbit spirals all the
way to the center. If such a tidal disruption is complete, and no bound
satellite is left, dynamical friction ceases to act. If the disruption is
only partial at this epoch, the surviving satellite fragment displays extensive tidal tails, leading and trailing its current position in the galactic halo.

While in $\Lambda$-Cold Dark Matter ($\Lambda$CDM) the interaction rate is expected to drop to the present-day epoch,
such disruption of satellites should still occur around normal spiral galaxies. 
The  fossil records of these merger events may be detected nowadays in the form of distinct coherent
 stellar structures in the outer regions of massive systems. The most spectacular cases of tidal debris
 are long, dynamically cold stellar streams from a disrupted dwarf satellite, which have wrapped around the host galaxy's disk and roughly trace the orbit 
of the progenitor satellite. The now well-studied
Sagittarius tidal stream surrounding the Milky Way (Majewski et al. 2003) and the
 giant stream in Andromeda galaxy (Ibata et al. 2007) are archetypes of these 
satellite galaxy merger 'fossils'  in the Local Group. 
They provide sound qualitative support for the scenario that tidally
disrupted dwarf
galaxies are important contributors to stellar halo formation in the Local Group spirals.

State-of-the-art, high-resolution numerical
simulations of galaxy formation, built within the $\Lambda$CMD context (e.g. Moore et al, 1999; Springel et al. 2008), can guide the quest for observational signatures of such star-streams (e.g. Bullock \& Johnston 2005; Johnston et al. 2008). Recent  simulations have demonstrated that the characteristics of substructure currently visible in the stellar halos are sensitive to the last (0-8 Gyr ago) merger histories of galaxies, a timescale that corresponds to the last few to tens of percent of mass accretion for a spiral galaxy like the Milky Way. While stellar streams in the Milky Way and Andromeda can be studied in detail, comparison with cosmological models is limited by 'cosmic variance'.
However, the current models imply that a survey of 50-100 parent galaxies reaching to a surface brightness
of 30 mag arcsec$^{-2}$ would reveal many tens of tidal features, perhaps nearly one 
detectable stream per galaxy (Johnston et al. 2008). However, a specific
comparison of these simulations with observations is missing because no suitable data sets exist.
Such a comparison, which could quantify the present sub-halo merger rate, is not only important as a test 
of $\Lambda$CDM models, but also as a more direct probe of how resilient disks are to minor mergers.

%% FIGURE 1
\begin{figure}[t]
  \begin{center}
   \epsfig{figure=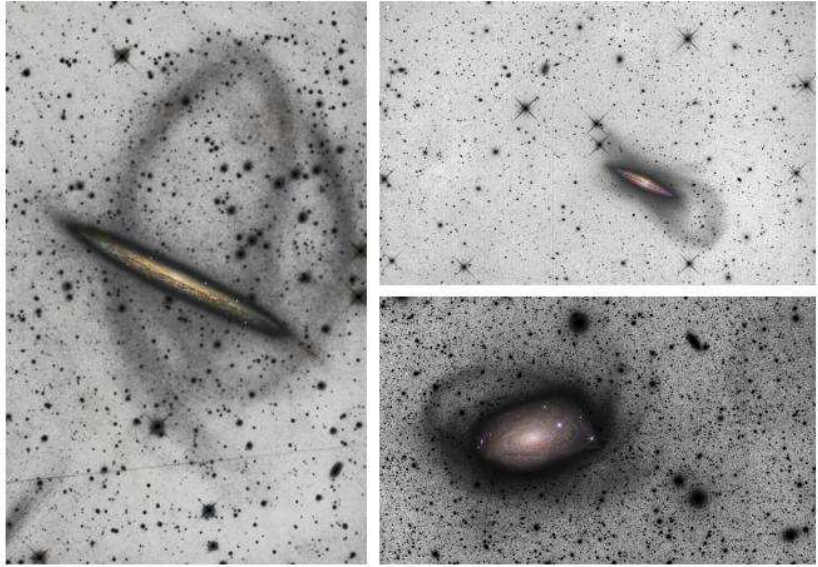, height=9cm, width=12cm, angle=0}
    \caption{ \footnotesize{{\it(left)} Deep image of the stellar tidal stream
around NGC 5907  obtained with the 0.5-meter Black Bird Observatory (BBO) telescope  (Martinez-Delgado et al 2008a). A  N-body model of this structure is shown in Fig. 2; {\it (right, top)} A low-galactic latitude stellar tidal stream
of NGC 4013, discovered by our team from deep images taken
with the BBO telescope; {\it (right, bottom)} Deep images
taking with a FSQ-106ED telescope of only 10cm aperture allowed
the discovery of a  giant tidal stream in the halo of the spiral
 galaxy Messier 63 (Martinez-Delgado et al. 2009, in preparation).
 A colour inset of the disk of each galaxy has been inserted with reference purpose.}}
    \label{fig_test}
  \end{center}
\end{figure}

\section{Stellar tidal streams in external galaxies}
\label{sec:2}

 {\it Galactic archeology} by looking at tidal remnants is a relatively new field of research that so far
has been primarily focused on the Local Group spiral galaxies. The first known tidal stream surrounding
the Milky Way (the Sagittarius tidal stream) was discovered less than one decade ago (Mateo et al. 1998; Mart\'\i nez-Delgado et al. 2001). In recent years, studies have focused on the formation and evolution of our Galaxy have been revolutionized by the first generation of wide-field, digital
imaging surveys. The resulting extensive photometric databases have provided,
for the first time, spectacular panoramic views of Milky Way tidal streams ( e.g. the {\it Field of Streams}: Belokurov et al. 2006) and have revealed the existence of  large stellar
sub-structures in the halo (Newberg et al. 2002; Juric et al 2008), which have been interpreted as observational evidence of our home Galaxy's hierarchical formation. The discovery of the Monoceros tidal stream (Yanny et al 2003) and
the possible Canis Major dwarf galaxy (Martin et al. 2004; Mart\'\i nez-Delgado
et al. 2005), located close to the Galactic plane, indicates that minor mergers might play a relevant role in the formation of the outer regions of spiral disks (Pe\~narrubia et al. 2006). A multitude of tidal streams, arcs, shells and other irregular structures that are possibly related to ancient merger events can be seen  in deep panoramic views of the Andromeda halo (Ibata et al. 2007).  These pictures show in detail the level of stellar sub-structure that might be present in the halos of nearby external spiral galaxies.

\vspace{0.2cm}

Our current understanding of the Local Group spirals has provided a couple of firsthand examples of individual minor mergers and their link to the current
state of massive galaxy building (Johnston et al. 2008). The search for analogues to these galactic fossils beyond
the Local Group is required not only to see whether the Milky Way and Andromeda galaxies are
'typical' with regard to substructure formation, but to  estimate the fractional contribution of accreted mass and the mass spectrum of accreted bodies in the life of these massive systems, an issue that remains  unresolved.

 Unfortunately, over the past decade only a few cases of confirmed stellar tidal streams have been detected
in spiral galaxies outside the Local Group (e.g. Pohlen et al. 2004 and references therein). The first cases of extragalactic tidal streams were reported a decade ago  by Malin \& Hardlin (1997). Using special contrast enhancement techniques on deep photographic plates, they were able to highlight  two possible tidal streams surrounding the galaxies  M83 and M104. Then, deep CCD images of the nearby, edge-on galaxy NGC 5907 by Shang et al. (1998), revealed  an elliptically-shaped
loop in the halo of this galaxy. This was the most
compelling example of a non-Local-Group tidal stream up to now.  More recently, very deep images (Martinez-Delgado et al. 2008a,b) have clearly revealed large scale, complex structures of arcing loops in the halos of several nearby galaxies (see Fig. 1). These detailed observations provide an elegant example of how a single, current epoch, low-mass satellite accretion can produce a very complex,
rosette-like structure of debris dispersed in the halo of its host galaxy.

%% FIGURE 2
\begin{figure}[t]
  \begin{center}
   \epsfig{figure=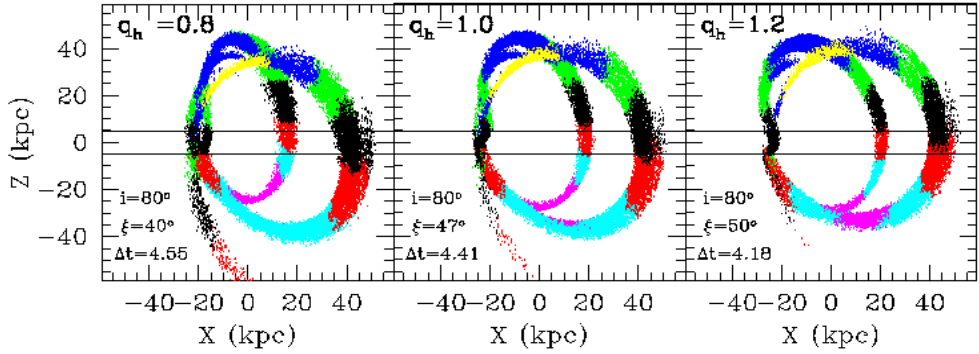, height=5cm, width=13cm, angle=0}
\caption{N-body model of the stellar stream detected in NGC
  5907. The satellite is realized as a King model with an initial mass, King
  core and tidal radii of $M=2\times10^8 M_\odot$, $r_c=0.39$ kpc and $r_t=2.7$ kpc, respectively (see Mart\'\i nez-Delgado et al. 2008a for details). For this particular model the orbital period  is $T_r=0.9$ Gyr. Each panels adopt a different halo minor-to-major axis-ratio ($q_h$). Different colors denote different ranges of projected radial velocities, namely yellow
(-210,-150) kms/s; blue (-150,-90) km/s; green (-90,-30) km/s; black (-30, +30) km/s; red
(+30,+90) kms; cyan (+90,+150) km/s and magenta (+150,+210) km/s. Note that the three
models show similar variations of the projected radial velocity along the stream. This result suggests that additional kinematical data may provide insufficient information to break the model degeneracies discussed in Martinez-Delgado et al. (2008a).}
    \label{model}
  \end{center}

\end{figure}

\section{A pilot survey of stellar tidal stream in nearby galaxies (2006-2008) }
We have initiated a pilot survey of stellar tidal streams in
a select number of nearby, edge-on spiral systems using modest (0.1-0.5-meter), robotic telescopes
 operating under very dark skies. The main results of this first observational
effort are given below.

\subsection{ The tidal stream of NGC 5907}
In summer 2006,  we re-observed the tidal stream of NGC5907 as a commissioning target to demonstrate  the sensitivity of our small aperture telescope for detecting extragalactic tidal streams.  Our deep observations showed for first time an interwoven, rosette-like structure of debris dispersed in the halo of this spiral galaxy (Fig.1; Martinez-Delgado et al. 2008a).  Its presence provides confirmation that these tidal remnants can survive several gigayears, as predicted by N-body simulations of tidally disrupted stellar systems around the
Milky Way (e.g., Law et al 2005; Pe\~narrubia et al. 2005).

Our N-body simulations (Fig.2) of the tidal disruption of a dwarf satellite by a disk galaxy and its dark halo potential suggest that
most of the tidal features observed in NGC\,5907 can be
explained by a single accretion event. Interestingly,  this  model finds that the stellar stream may be
relatively old with the fainter, outer loop material  becoming unbound at
least 3.6 Gyrs ago. The stellar stream around NGC 5907 may therefore represent one
of the most ancient tidal debris ever reported in the halo of a
spiral galaxy. It also confirms
that spiral galaxy halos in the Local Universe still contain a
significant number of galactic fossils from their hierarchical
formation and that they can be detected with modest instruments.

%% FIGURE 4
\begin{figure}[t]
  \begin{center}
   \epsfig{figure=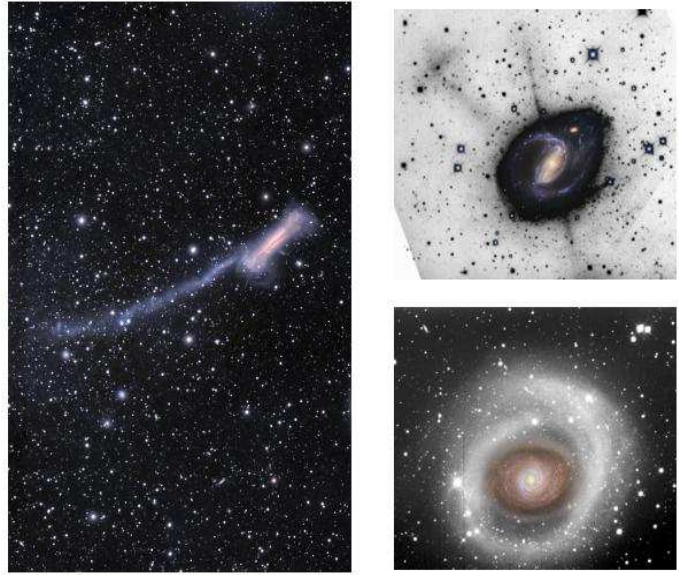, height=9cm, width=11cm, angle=0}
    \caption{ \footnotesize{Diffuse light, giant structures detected in the outskirts of several nearby galaxies in our pilot survey: {\it (left)} Panoramic
view of the halo of NGC 3628 obtained with a  10-cm telescope.
In addition to the known giant filament (Chromey et al. 1998), our deep images reveal for first time new arc-like features in the edge of its perturbed disk, which
suggest that this galaxy could be suffering a tidal encounter with
a dwarf satellite; {\it (bottom, right)} A striking, very faint ring structure in M94 detected with the BBO telescope (Trujillo et al. 2009, in prep.), consistent with being the optical counterpart of the ultraviolet extended disk discovered from
GALEX observations (Gil de Paz et al. 2007);
{\it (top, right)} First wide-field CCD panoramic image of the mysterious jet-like features in the halo of the active galaxy
NGC 1097, previous reported in photographic plates (Arp 1976). The tidal origin of
these features is still controversial (Higdon \& Wallin 2003).}}
    \label{zoo}
  \end{center}
\end{figure}

 \subsection{ Discovery of stellar tidal streams in warped disk galaxies }

Promising galaxies in the hunt for extragalactic tidal streams are those
that display outstanding asymmetries in optical or HI images. It has been long
suggested that these perturbations are a result of gravitational interaction with nearby companions. The most striking case is NGC 4013, an isolated spiral
galaxy famous for having one of the most prominent HI warps detected so far (Bottema et al. 1987). Our deep images of this galaxy (Mart\'\i nez-Delgado et al. 2008b) revealed a faint  loop-like structure that appears to be part of a gigantic, low-inclination stellar tidal stream (Fig. 1).  Although its true three dimensional geometry is unknown,  the sky- projected morphology of this structure displays a remarkable resemblance to  the theoretical predictions for an edge-on view of the Milky Way's Monoceros tidal stream (Pe\~narrubia et al. 2005). This suggests that the progenitor system (whose current position and final fate remains unknown) may have been a galaxy with an initial mass $\sim 10^8 M_\odot$ moving on a low-inclination ($\simeq 25^\circ$), nearly-circular orbit. Using this model as a template, the tidal stream may be approximately $\sim 3$ Gyr of age.

We have also discovered a stellar tidal stream in the halo of the nearby spiral Messier 63 (Martinez-Delgado et al. 2009, in preparation). Our data, collected from different
telescopes, reveals an enormous, arc-like structure around this galaxy's disk extending  $\sim$29 kpc from its center (Fig.1),apparently tilted with
respect to its strong  gaseous warp. This  unassailable indication of a past merger event provides an additional example of apparently isolated galaxies with significantly warped gaseous disks that also show evidence indicating the ongoing tidal disruption of a dwarf companion. 

These results highlight the fact that disks that appear to be undisturbed in high surface brightness optical images but warped in HI  maps can reveal complex signatures of recent accretion events when viewed in deep optical surveys. Additionally, with the
growing number of examples of spirals (NGC 5907, NGC 4013, M63) showing a connection between warped disks and evidence of mergers, the origin of galactic warps via interactions with minor mergers is a possibility worth further investigation. 

\subsection{  Diffuse light structures in nearby spiral galaxies}

Over the last two years, we have also obtained wide field, follow-up images of  a dozen familiar,  nearby spiral galaxies that are widely known to have diffuse light features in their outer regions. Although our current data for them is not yet deep enough yet (with an estimated surface brightness limit of $\sim$ 28 mag/arcsec$^{2}$), our  panoramic views of their halos  have revealed  their fossil structures in detail. The most conspicuous cases detected during this observational effort are shown in Fig. 3. In addition, we have also discovered striking  stellar tidal streams of different morphological types in several  neighboring galaxies (Martinez-Delgado et al. 2009, in preparation). Thus, our deep images are very effective in revealing a plethora of  very faint morphological perturbations  and dynamical features in the external regions of nearby galactic disks, presumably all 
signposts of minor gravitational interactions.

\section{Future work} 

 The promising results of our foray into a more systematic look for tidal streams in the nearby universe encourage a more aggressive attention to this new way of understanding galaxy formation. The overall goal of this  project is to conduct the first systematic survey of stellar tidal streams
for $\sim 50$ nearby spiral galaxies (D$<$ 15 Mpc) to a (large area) surface brightness sensitivity of
$\sim 30$ mag/sqasec, in order to obtain the first comprehensive census of such
structures in the Local Volume. Based on theoretical predictions on stream counts (one stream expected per galaxy; see Sec.1), we expect this observational effort to  yield a tidal stream sample with sufficient statistical significance to undertake a direct comparison  with high resolution cosmological simulations of hierarchical galaxy formation and satellite dark matter halo dynamics.

 A number of astrophysical problems can be tackled with the data produced from this survey and covering a large variety of classical topics (stellar populations, formation of galactic disk, galactic dynamics, globular cluster formation,dark matter halo flattening; cosmology; see a discussion in Mart\'\i nez-Delgado et al. 2008a). The survey will result in estimates of the incidence, size/geometry and stellar luminosity/mass
distribution of such streams. The results of the project will provide a direct and stringent test of
hierarchical structure formation on this scale, will constrain the present-epoch (minor) interaction rate and probe the minor-merger resilience of stellar disks.

\begin{acknowledgement}

The authors gratefully thank the following astrophotographers whose collaboration helped make our initial survey possible: Ray Gralak (Fig. 2, M63), Steve Mandel (Fig. 3, NGC 3628), Kenneth Crawford and Mischa Schirmer.

\end{acknowledgement}

\end{document}